\documentclass[11pt]{elsarticle}
\usepackage{graphics}
\usepackage{graphicx}
\usepackage{epsfig}
\usepackage{amsmath}
\usepackage{multirow}
\usepackage{amssymb}
\usepackage{ragged2e}
\usepackage{setspace}

\newcommand*{\beq}{\begin{equation}}
\newcommand*{\eeq}{\end{equation}}
\newcommand*{\btab}{\begin{table}}
\newcommand*{\etab}{\end{table}}
\newcommand{\Teleia}{.}
\newcommand{\Comma}{,}
\newcommand{\Deff}{D}
\journal{Journal of Theoretical Biology}

\begin{document}

\title{Spatial dynamics of airborne infectious diseases}

\author[jrc]{M.�Robinson\corref{cor1}}
\ead{marguerite.robinson@jrc.ec.europa.eu}
\author[jrc,ger]{N.I.�Stilianakis}
\ead{nikolaos.stilianakis@jrc.ec.europa.eu}
\author[jrc]{Y.�Drossinos}
\ead{ioannis.drossinos@jrc.ec.europa.eu}

\cortext[cor1]{Corresponding author,Tel: +39 0332 786343}
\address[jrc]{Joint Research Centre, European Commission, I-21027 Ispra (VA), Italy}
\address[ger]{Department of Biometry and Epidemiology, University of Erlangen-Nuremberg, Erlangen, Germany}

\begin{abstract}
Disease outbreaks, such as those of Severe Acute Respiratory Syndrome in 2003 and the 2009 pandemic A(H1N1) influenza, have highlighted the potential for airborne transmission in indoor environments. Respirable pathogen-carrying droplets provide a vector for the spatial spread of infection with droplet transport determined by diffusive and convective processes. An epidemiological model describing the spatial dynamics of disease transmission is presented. The effects of an ambient airflow, as an infection control, are incorporated leading to a delay equation, with droplet density dependent on the infectious density at a previous time. It is found that small droplets ($\sim 0.4\ \mu$m) generate a negligible infectious force due to the small viral load and the associated duration they require to transmit infection. In contrast, larger droplets ($\sim 4\ \mu$m) can lead to an infectious wave propagating through a fully susceptible population or a secondary infection outbreak for a localised susceptible population. Droplet diffusion is found to be an inefficient mode of droplet transport leading to minimal spatial spread of infection. A threshold air velocity is derived, above which disease transmission is impaired even when the basic reproduction number $R_{0}$ exceeds unity.
\end{abstract}

\maketitle

\section{\label{sec1}Introduction}
The temporal dynamics of disease transmission is well documented
\cite{anderson}, however, the spatial spread of infection and
the mechanisms driving it are less well understood. Spatial dynamics
of respiratory diseases has received much attention in recent years
following the rapid global spread of both Severe Acute Respiratory Syndrome (SARS) \cite{hufnagel} and
pandemic influenza (2009) A(H1N1) \cite{balcan}. Large-scale geographic models are typically implemented by
superimposing a transportation network on local infection dynamics
\cite{keelinga,colizza}. Smaller
scale models often implement reaction-diffusion equations to
describe random movements within populations \cite{noble,murraya}.
Such models have been shown to exhibit
traveling-wave solutions whose existence depends on the basic
reproduction number $R_{0}$ \cite{kallen,mendez}. However, few models address the mode of disease transmission
that underlies the spatial dynamics.

For some respiratory infections, such as influenza, three modes of transmission have been
identified: airborne, droplet and contact transmission \cite{weber}. All three modes result from the generation of
respiratory droplets by an infected person during an expiratory
event (e.g. coughing, sneezing). Droplet and contact transmission
require relatively close contact between the infected and susceptible
individuals for efficient disease transmission. Therefore, spatial
spread via these routes of transmission must be driven by human
movement. In contrast, the airborne spread of a pathogen may be attributed
to the movement of both people and fine aerosol droplets
suspended in the air and their associated airborne residence time and pathogen load. Furthermore, the airborne route is the primary mode of transmission for other respiratory diseases, such as tuberculosis, and its contribution to the spatial spread of infection is of paramount importance.

The relative importance of the three modes of transmission is
difficult to quantify; however, recent experience with SARS and
influenza outbreaks has highlighted the potential for airborne
transmission in indoor environments. Evidence of the airborne
transmission of respiratory infections has been documented in
hospital ward settings \cite{wong}, housing
complexes \cite{yu} and on board commercial airliners \cite{moser}. In particular, the importance of
indoor airflows has been established \cite{li}, with large
air flow rates resulting in lower disease
transmission \cite{nielsen}. However, the threshold air velocity above
which disease transmission is impaired is as yet unknown. A
comprehensive knowledge and understanding of the mechanisms driving
such transmission is vital for the implementation of adequate
infection controls in indoor public environments such as schools,
hospitals and long-term care facilities.

Airborne transmission is mediated by fine aerosol droplets small
enough to remain suspended in air for prolonged periods, and large enough
to contain non-negligible pathogen load. The
standard epidemiological models operate on the
assumption that contact between susceptible and infected persons is
necessary for disease transmission \cite{keelingb}.
However, the pathogen-carrying droplets emitted by an infected individual during an
expiratory event are the disease vector
and the standard models
should be adjusted to reflect this. Such models have been used to
describe the spatial spread of fungal spores over a vineyard \cite{burie} and to model disease spread following the
point release of an infectious agent \cite{reluga}.

A zero-dimensional model for the temporal development of an epidemic driven by expiratory
droplets (in particular, respirable droplets)
was developed in \cite{Niko}. The model is built on the concept
of an infectious cloud surrounding each infected individual, an idea also considered in \cite{eichenwald}.
In this work we investigate the airborne spread of an infection through a closed spatial environment within which the human population is confined for a prolonged period of time. For example, such outbreaks have been documented in public settings including prisons \cite{awofese,gomez}, boarding schools \cite{rose}, long-term care facilities \cite{dharan} and on board large cruise ships \cite{miller,ward}. In the absence of intervention measures, such outbreaks can persist for many weeks. We extend the zero-dimensional model to include the spatial dynamics of airborne-droplet transmission (also known as aerosol transmission)
and investigate how an ambient airflow can
influence disease spread. Accordingly, we neglect transmission by
droplet spray, a close-contact transmission mode that occurs via direct
deposition onto a susceptible's mucous membranes, and by (physical) contact
transmission. Furthermore, we consider transmission only by respirable expiratory
droplets, droplets whose post-evaporation diameter is less than $10\ \mu$m,
neglecting transmission by inspirable droplets, droplets whose post-evaporation
diameter is between $10$ and $100\ \mu$m. We follow \cite{nicas} and take the droplet post-evaporation diameter to be half the pre-evaporation diameter. Inspirable droplets contribute to disease
transmission by inhalation almost immediately after generation (e.g., during the first
breath) as they are considerably larger than respirable droplets and they gravitationally
settle very fast. As in the case of droplet spray, transmission by inspirable
droplets occurs only at close contact.

\section{\label{sec2}A one-dimensional spatial model for airborne transmission}
Consider a population of susceptible, infected and recovered individuals. Let $S(x,t)$ be
the density of susceptibles, $I(x,t)$ the density of infected and $R(x,t)$ the density of recovered
individuals, with $N(x,t)=S+I+R$ being the total population density and $n(t_{0})=\int_{x}{N(x,t_{0})}dx$
representing  the total number of people in the spatial domain at any time $t_{0}\geqslant0$.
Henceforth, all densities refer to spatial densities, unless otherwise noted,
and airborne droplets refer to respirable droplets.
It is assumed that infected individuals
continuously generate a cloud of pathogen carrying aerosol droplets and we
let $D(x,t)$ be the (number) density of active droplets, which we define as droplets that are both airborne and have a nonzero pathogen load.

The zero-dimensional model, which constitutes the basis of the one-spatial dimension model, is derived in Stilianakis and Drossinos \cite{Niko}. A general one-dimensional evolution equation to model the spatial spread of disease due to the continuous motion of people and droplets takes the form
\begin{displaymath}
\frac{\partial C_{i}}{\partial t}=-\frac{\partial Q_{i}}{\partial x}+\Psi_{i}(x,t) \Comma
\end{displaymath}
for each species $i\in[S,I,R,D]$, where $C_{i}(x,t)$ represents the density of species $i$, the flux $Q_{i}(x,t)$ is the rate at which species $i$ passes the point $x$ at time $t$ and $\Psi_{i}(x,t)$ is a source term representing the creation (or destruction) of species $i$. The density of
susceptibles will decrease through contact with pathogen-carrying (respirable airborne) droplets and subsequent infection. The density of infected people will increase accordingly. Furthermore, it will decrease at the rate that individuals recover $\mu_i$, where $1/\mu_i$ is the disease infectivity period. Thus,
\begin{displaymath}
\Psi_{S}=-\frac{\beta_{d}}{N}\Deff S,\quad\Psi_{I}=\frac{\beta_{d}}{N}\Deff S-\mu_i I,\quad \Psi_{R}=\mu_i I \Comma
\end{displaymath}
where $\beta_{d}$ is the transmission rate per droplet of diameter $d$. A detailed derivation of the transmission terms is provided in \cite{Niko}. The dynamics of the airborne droplets is determined by generation and annihilation processes.
The droplet density at any point is proportional to the density of infected individuals and increases at the rate $\kappa_{d}$ that pathogen-loaded droplets are shed. Moreover, the active droplet density will decrease as droplets are removed through gravitational settling and inhalation (by the person who generated it or another population member) and via pathogen inactivation. Thus,
\begin{displaymath}
\Psi_{D}=\kappa_{d} I-\alpha_{d} \Deff \Comma
\end{displaymath}
where $\alpha_{d}$ is the droplet removal rate and $1/\alpha_d$ the droplet infectivity period.

We assume that human movement can be modeled as a diffusive process \cite{noble,keelingb,murrayb} and the associated flux is given by Fick's law as
\begin{displaymath}
Q_{j}=-D_{p}\frac{\partial C_{j}}{\partial x},\quad\textrm{for}\quad j=S,I,R \Comma
\end{displaymath}
where the minus sign is interpreted as the tendency of people to move from high density areas to low density areas and $D_{p}$ is the diffusivity of the human population. For real world situations, movement of the  general population might also be modeled by including a convective term, whereby human motion would be faster and in a specified direction. However, movement in such facilities as prisons and long-term care facilities is more restricted and sporadic and a diffusive flux better models such a scenario. In addition, restricting movement to a diffusive process is more convenient to investigate if droplets can drive the transmission process. 

The droplet flux $Q_D$ depends on environmental characteristics. Droplets are generated through expiratory events (e.g. coughing, sneezing) with an initial velocity. Average expiration velocities are $11.7$ ms$^{-1}$ and $3.9$ ms$^{-1}$ for coughing and speaking respectively
\cite{chao}. For respirable droplets, the droplet relaxation time, the time required to adjust the droplet velocity to a new condition of forces, is of the order  $\sim 10^{-7} - 10^{-4}$ s. Therefore, the droplet velocity rapidly tends to the carrier-gas (air) velocity. The exhaled air flow has been modelled as a continuous turbulent round jet \cite{wang,parienta}. For a steady-state turbulent jet, small droplets, which follow the instantaneous fluid streamlines, may reach relatively large distances from the source (more than $8$ m after $100$ s). However, it was argued in \cite{zhu},
where the air flow within a calm room was calculated via a Computational Fluid Dynamics software, that droplets smaller than $30\mu$m diameter would disperse within the room without significant influence of gravity or inertia. Accordingly, we consider that the initial velocity of the exhaled respirable droplets is the underlying fluid velocity, namely either the ambient air velocity, or zero in its absence. For convenience, we henceforth refer to the ambient airflow as the ventilation. This airflow could be naturally induced (e.g. a draft through an open window) or be the result of an artificial indoor ventilation system.

It follows that, in an enclosed space with no ventilation, airborne droplets will be transported by molecular  diffusion alone and $Q_{D}=-D_{d}\ \partial\Deff/\partial x$, where $D_{d}$ is the molecular droplet diffusivity. Molecular (Brownian) diffusivity may be safely ignored for droplets larger than $0.5\mu$m diameter. We retain it in the formal derivation of the flux equations to allow the simulation of nanodroplets, to allow for turbulent droplet diffusion in the presence of an external turbulent flow, and to render the numerical solution of the parabolic differential
equations stable. In a ventilated environment airborne droplets will also be convected by the ambient airflow. The appropriate flux is $Q_{D} = v \Deff -D_{d}\ \partial\Deff/\partial x$, where $D_d$ is the turbulent diffusivity, and $v$ the average, constant air ventilation velocity. The one-dimensional droplet flux is an idealised approximation. In reality, the air flow is unlikely to be at a constant unidirectional velocity, and ventilation may simply mix droplet and air particles around within the domain. A higher dimensional spatial model is required to describe such complex flow characteristics. However, the one-dimensional model can be a good approximation for environments where air is predominantly in one direction, for example a draft of air blowing through a room.\\
Assuming the diffusion coefficients of both people and droplets are constant the model is
\begin{align}
\frac{\partial S}{\partial t}&=-\beta_{d}\frac{\Deff S}{N}+D_{p}\frac{\partial^{2} S}{\partial x^{2}} \Comma \label{seqn}\\
\frac{\partial I}{\partial t}&=\beta_{d}\frac{\Deff S}{N}-\mu_i I+D_{p}\frac{\partial^{2} I}{\partial x^{2}} \Comma \label{ieqn}\\
\frac{\partial N}{\partial t}&=D_{p}\frac{\partial^{2} N}{\partial x^{2}} \Comma \label{neqn}\\
\frac{\partial \Deff}{\partial t}&=\kappa_{d} I-\alpha_{d} \Deff -v\frac{\partial \Deff}{\partial x}
+D_{d}\frac{\partial^{2} \Deff}{\partial x^{2}} \Comma
\label{deqn}
\end{align}
and the density of recovered individuals is obtained from $R=N-(S+I)$. Equations~(\ref{seqn}-\ref{deqn}) reduce to the model equations proposed in \cite{Niko} for
infectious disease transmission by respirable droplets without contact transmission with zero ventilation velocity and diffusion coefficients.

We consider the spread of a pathogen in the domain $0\leqslant x\leqslant l$. We assume that people are confined to the interval $[0,l]$ and prescribe zero-flux conditions at the boundaries (the subscript $x$ denotes partial differentiation with respect to $x$),
\begin{displaymath}
S_{x}(0,t)=S_{x}(l,t)=0 \Comma \quad\quad I_{x}(0,t)=I_{x}(l,t)=0 \Comma \quad\quad N_{x}(0,t)=N_{x}(l,t)=0.
\end{displaymath}
The boundary conditions for droplets would be expected to depend on the type of ventilation present. We assume that droplets are removed from the system when they reach the end of the domain (physically this could be, for example, through a wall vent or an open window). One option is then to place an artificial boundary at the end of the domain $x=l$ so that droplets effectively travel through the boundary unhindered. This is referred to as a transparent boundary condition. For convection-diffusion problems such a condition would take the form \cite{wesseling},
\begin{displaymath}
D_t(l,t)+vD_x(l,t)=0.
\end{displaymath}   
However, for a small diffusion coefficient, the boundaries are sufficiently far from the region of interest that, for the timescales of interest, droplets never reach them by diffusive processes alone. Therefore, it is sufficient to assume that droplets can be transported out of the domain by convection alone and we set the diffusive flux at the boundaries to zero,
\begin{displaymath}
D_{x}(0,t)=D_{x}(l,t)=0 \Teleia
\end{displaymath}
In our simulations we compared the use of the transparent and diffusive boundary conditions and found that, as expected, solutions were identical. Therefore, we present only those performed with the zero diffusive flux condition.\\

The prescribed boundary conditions prevent the entrance or exit of people. Therefore, people are free to move about the domain but cannot leave it. Consequently, the total number of
people in the domain will be constant for all time
\begin{displaymath}
n(t_{0})=\int_{0}^{l}N(x,t_{0})\ dx = n_{0},\ \forall\  t_{0}\geq 0 \Teleia
\end{displaymath}
Initial conditions must also be prescribed to distribute people throughout
the domain at $t=0$. Since the total number of people in the
domain is constant for all time, $n_{0}$ is determined from the initial distribution
\begin{equation}\label{constraint}
n_{0}=\int_{0}^{l}N(x,0)\ dx =\int_{0}^{l} \Big [ S(x,0)+I(x,0) \Big ]\ dx \Comma
\end{equation}
where we assume the initial number of recovered (immune) people is
identically zero, $R(x,0)=0$. Initially, we assume
that no droplets are present and are generated for $t>0$
by the infected population. We consider two possibilities for the initial distribution of the
human population. The total population can be uniformly distributed throughout the domain
\cite{noble}, which we henceforth refer to as a (spatially) \textit{homogeneous} initial condition.
This corresponds to a constant initial total population density
\begin{displaymath}
N_{\textrm{homo}}(x,0)=\frac{n_{0}}{l} \Teleia
\end{displaymath}
In this scenario, $N$ will be constant for all time, with $N(x,t)=n_{0}/l$. Alternatively, people can
be randomly distributed throughout the domain yielding a (spatially) \textit{heterogeneous} initial condition.
The latter case presents an opportunity to investigate whether droplets generated by infected individuals
at one point in the domain can infect susceptible individuals at other locations,
i.e., whether infection can occur without direct physical contact of a susceptible
with an infected individual.

\subsection{\label{sec2sub1}Scaling and non-dimensionalisation}
We scale all population densities with the uniform density $n_0/l$.
The droplet scale is chosen by balancing droplet generation and removal processes such that, the number
of droplets is approximately proportional to the number of infected
people and the constant of proportionality is the ratio of the
droplet generation and removal rates. We scale $x$ with the length of the domain and choose the droplet removal time as the characteristic time scale, since this represents the time the droplet is airborne and capable of causing infection. Therefore, we scale
\begin{equation}\label{scales}
S,I,N\sim\frac{n_{0}}{l},\quad\quad
\Deff \sim\frac{\kappa_{d}}{\alpha_{d}}\frac{n_{0}}{l}, \quad\quad x\sim l,\quad\quad t\sim\frac{1}{\alpha_{d}} \Teleia
\end{equation}
The dimensionless equations are
\begin{align}
\frac{\partial S}{\partial t}&=-\lambda R_{0}\frac{DS}{N}+\eta_{p}\frac{\partial^{2} S}{\partial x^{2}} \Comma \label{Susceptibleeqn}\\
\frac{\partial I}{\partial t}&=\lambda R_{0}\frac{DS}{N}-\lambda I +\eta_{p}\frac{\partial^{2} I}{\partial x^{2}} \Comma \label{Infectedeqn}\\
\frac{\partial N}{\partial t}&=\eta_{p}\frac{\partial^{2} N}{\partial x^{2}} \Comma \label{Populationeqn}\\
\frac{\partial D}{\partial
t}&=I-D-\nu\frac{\partial D}{\partial
x}+\eta_{d}\frac{\partial^{2} D}{\partial
x^{2}} \Comma \label{Dropleteqn}
\end{align}
where
\begin{displaymath}
R_0=\frac{\beta_d \kappa_d}{\alpha_d \mu_i},\quad\quad \lambda=\frac{\mu_i}{\alpha_d}, \quad\quad \nu=\frac{v}{\alpha_d l},\quad\quad \eta_{d,p}=\frac{D_{d,p}}{\alpha_{d}l^{2}} \Teleia
\end{displaymath}
The dimensionless parameters and the characteristic time scales of the model are summarized in Table~\ref{table:Scales}. The parameter $R_{0}$ is the basic reproduction number, which describes the spread of disease through a completely susceptible population in the initial stages of an outbreak. $\lambda$ is the ratio of the droplet removal time scale $\tau_r$ to the disease infectivity time $\tau_i$ and represents the fraction of the total disease infectivity period for which a droplet is capable of causing infection. The dimensionless coefficient of the convection term $\nu=\tau_r/\tau_c$ represents the ratio of the droplet-removal time to the convection time. Similarly, the dimensionless number $\eta_d=\tau_r/\tau_d$ represents the ratio of the removal time to the diffusion time. Boundary and initial conditions in dimensionless form are given by
\begin{displaymath}
\begin{array}{cl}
S_{x}(0,t)=S_{x}(1,t)=0 \Comma \quad & \quad S(x,0)= S_0(x) \Comma \\
I_{x}(0,t)=I_{x}(1,t)=0 \Comma \quad & \quad I(x,0)=I_0(x) \Comma \\
N_{x}(0,t)=N_{x}(1,t)=0 \Comma \quad & \quad N(x,0)=S_0(x)+I_0(x) \Comma \\
D_{x}(0,t)=D_{x}(1,t)=0 \Comma \quad & \quad D(x,0)=0 \Comma \\
\end{array}
\end{displaymath}
where the functions $S_0(x)$ and $I_0(x)$ are prescribed initial population-density distributions.
The dimensionless form of the initial population density, Eq.~(\ref{constraint}), is
\begin{displaymath}
\int_{0}^{1}N(x,0)\ dx = \int_{0}^{1} \Big [ S(x,0)+I(x,0) \Big ]\ dx = 1 \Teleia
\end{displaymath}
For the homogeneous case, with constant initial population density $N_{\textrm{homo}}(x,0)=1$, this implies
\begin{displaymath}
S(x,0)+I(x,0)=1 \Teleia
\end{displaymath}

\section{\label{sec3}Model parameters for an influenza outbreak}
We apply the one-dimensional model to numerically study the spatial and temporal dynamics of a model for an influenza epidemic. Most of the required parameters are taken from \cite{Niko}; the parameters specific to the spatial dynamics and droplet size are discussed in the following, and parameter values used in the numerical simulations are summarized in Table~\ref{table:Parameters}.

The effect of droplet size on disease spread is investigated by choosing two characteristic respirable-droplet sizes of post-evaporative diameters
$d_{1}=4$ $\mu$m and $d_{2}=0.4$ $\mu$m. As mentioned earlier, the post-evaporative diameter is taken to be half the emitted pre-evaporative diameter. Droplet generation rates $\kappa_{d}$ are based on the number of pathogen loaded droplets emitted during a cough. Generations rates
per cough are taken as $\kappa_{d_{1}}=160$ day$^{-1}$ \cite{nicas} and $\kappa_{d_{2}}=240$ day$^{-1}$ \cite{morawska}, the latter based on a cough volume of $400$ cm$^{3}$. The daily generation rates are obtained by considering a 200-fold increase for a sneeze \cite{nicas}, and a total of 11 sneezes and 360 coughs per day \cite{atkinson}.

In the model presented here, the infectious agent is not the droplet but the pathogens it carries. Therefore, the transmission rate per droplet $\beta_d$ will depend the transmission rate per pathogen, $\beta_d=\beta_pq_dN_p^0$, where $q_d$ is the probability of deposition in the human respiratory tract and $N_p^0(d)$ is the number of pathogens in a droplet of diameter $d$. In turn, the transmission rate per pathogen $\beta_p$ is determined from the contact rate $c_d$ of a susceptible with a droplet and the probability $p_d$ that such a contact will result in successful transmission $\beta_p=c_dp_d$. In order to derive the contact rate with a droplet we assume that each infected person is surrounded by a droplet cloud with volume $V_{cl}$. It is further assumed that a susceptible individual comes in contact with a droplet through breathing during an encounter with this droplet cloud. If the average breathing rate is $B$ and $\tau_{ct}$ is a characteristic time of breathing during the encounter then the contact rate $c_d$ can be expressed as $c_d=c\frac{B}{V_{cl}}\tau_{ct}$, where $c$ is the average number of total contacts
a susceptible individual has per unit time. The transmission rate per droplet is thus
\begin{displaymath}
\beta_{d}=c\frac{B}{V_{cl}}\tau_{ct}p_dq_dN_p^0,
\end{displaymath}
and the number of pathogens per droplet can be determined by $N_p^0=V_d\rho_p$, where $V_d$ is the volume of the (spherical) pre-evaporative droplet and $\rho_p$ is the pathogen concentration of the lung fluid. Using the parameters of Table~\ref{table:Parameters}, the transmission rate per pathogen
is calculated as $0.028$ per day and the transmission rate per droplet thus evaluates to $\beta_{d_{1}}=2.45\times10^{-5}$ per day and $\beta_{d_{2}}=5.57\times10^{-9}$ per day.

The droplet removal rate is determined by three distinct processes: gravitational settling of the droplet, inactivation of the pathogen load (which effectively removes the droplet) and the inhalation of the droplet by population members. Following Stilianakis and Drossinos \cite{Niko} the removal rate can be expressed as
\begin{equation}
\alpha_d=\theta_d+\mu_p+\left(1+c\tau_{ct}\right)\frac{B}{V_{cl}}q_d \Comma
\label{DropletRemovalRate}
\end{equation}
where $\theta_d$ is the gravitational settling rate of the droplet and $\mu_p$ is the inactivation rate of airborne pathogens. Gravitational settling rates are size dependent and we take $\theta_{d_{1}}=28.8$ per day and $\theta_{d_{2}}=0.39$ per day \cite{drossinos}. This implies that, in the absence of other removal processes, a droplet with diameter $d_{1}=4$ $\mu$m will settle under gravity in a tranquil environment in approximately 50 minutes. Decreasing the droplet diameter by an order of magnitude results in the droplet remaining airborne for approximately 2.56 days. The pathogen inactivation rate $\mu_p$, assumed to be independent of size, is taken to be $8.64$ per day \cite{hemmes} and a droplet is effectively removed
through pathogen inactivation in under 3 hours. Therefore, pathogen inactivation is a crucial process for the removal of smaller
droplets that could, theoretically, take days to settle under gravity alone. Taking both gravitational settling and pathogen inactivation processes into account, droplets remain airborne and infectious for $38.5$ minutes ($d_{1}$) and $2.66$ hours ($d_{2}$). The inclusion of removal through inhalation, the last term on the right-hand-side of Eq.~(\ref{DropletRemovalRate}), has negligible influence on these times (removal times of $35$ mins or $2.46$ hrs for $d=d_{1}$ or $d=d_{2}$ respectively) and we approximate the droplet removal rate by $\alpha_{d}=\theta_{d}+\mu_{p}$.

Droplet diffusivity will depend on the presence or absence of ventilation. In an unventilated environment, with $v\approx0$ m
s$^{-1}$ the molecular diffusivity of droplets is calculated as $D_{d_{1}}=6.2\times10^{-12}$ m$^{2}$ s$^{-1}$ and
$D_{d_{2}}=8.34\times10^{-11}$ m$^{2}$ s$^{-1}$ \cite{drossinos}. For an air-conditioned environment with standard wall-mounted
air-conditioners a typical airflow velocity is $v=0.2$ m s$^{-1}$ \cite{zhu} and we take this to be constant
throughout the entire domain. Under such conditions the flow will invariably be turbulent and the droplet diffusivity will be several
orders of magnitude larger. We estimate the turbulent diffusivity of both droplet classes to be $D^{tur}_{d}=10^{-3}$ m$^{2}$ s$^{-1}$.

Human diffusivity, motion, can be crudely estimated from $D_{p}\approx x^{2}/t$, where $x$ is a characteristic
distance traveled in a time $t$. For the spatial spread of a disease through a geographically open population typical distances traveled per hour are in the range $18-42$ m \cite{noble,bertuzzo,lou}. We expect that movement would be more restricted in a closed environment, for example a prison or long-term care facility, and we estimate that people diffuse approximately 10 meters per hour yielding $D_{p}\approx 10^{-5}$ m$^{2}$ s$^{-1}$. Intuitively, we estimate the spatial size of such environments to be of the order $10^3$ meters and and fix our domain length at $l=2000$ m.

\section{\label{sec4}Droplet dynamics}

\subsection{Characteristic time scales}
The time scale of disease transmission
$\tau_{t}=\alpha_{d}/\beta_{d}\kappa_{d}$ is determined from
droplet properties and is thus dependent on droplet size. We
estimate $\tau_{t}(d_{1})\approx3.72$ days and
$\tau_{t}(d_{2})\approx2640$ days. Clearly, the time required by the
smaller droplet to transmit disease will result in minimal disease
transmission for the duration of the infectivity period (approximately $5$ days for influenza).

Droplet dynamics are greatly influenced by their (post-evaporative)
diameter as this determines the residence time in air of an
individual droplet, its infectivity properties and its pathogen load. The dynamics are
described by three different timescales (Table \ref{table:Scales}), two of which, $\tau_{r}$ (droplet removal time scale)
and $\tau_{d}$ (droplet diffusion time scale), are dependent on droplet size.
The convective timescale $\tau_{c}$ represents the time it takes for the ambient airflow to
carry the droplet the entire length of the domain. If $l$ is of the
order of $10^{1}-10^{3}$ m then typical convection times are
$\tau_c=O(10^{2}-10^{4})$ seconds and, clearly, convection is a fast process relative to the disease dynamics with $\tau_c\ll\tau_t$. Intuitively, this implies that lower ventilation velocities, which increase the convection time $\tau_c$, result in greater disease transmission.
The droplet-removal (droplet infectivity) time scale is the
inverse of the size-dependent droplet removal rate $\alpha_{d}$ with
$\tau_{r}(d_{1})\approx 38.46$ minutes and $\tau_{r}(d_{2})\approx2.66$ hours.
Thus, since $\tau_{r}(d_{1})<\tau_{r}(d_{2})$ we have that $\nu_{d_{1}}<\nu_{d_{2}}$ and
convection effects, for a constant air-ventilation velocity,
are stronger for smaller droplets since they remain airborne longer
and can be carried further from their point of origin. Furthermore,
$\nu$ is inversely proportional to the length $l$ and convection
will thus exert greater influence on droplet dynamics over shorter
domains as droplets will be rapidly transported the entire length of
the domain.

The droplet diffusion time scales depend on the presence or absence of an
ambient airflow. For an unventilated environment, typical Brownian diffusion
time scales
will be $\tau_{d}(d_{1})\sim10^{9}-10^{13}$ days and
$\tau_{d}(d_{2})\sim10^{8}-10^{12}$ days for $l\sim 10^{1}-10^{3}\
\textrm{m}$. In a ventilated environment, we estimate the time scale
of turbulent diffusion to be $\tau^{tur}_{d}\sim1-10^{4}$ days. Thus,
for the short term spread of infection, since droplets are only
airborne for $O(1 \textrm{ hour} )$ and $\tau_r\ll\tau_d$, diffusion is an insufficient mechanism to transport droplets throughout the domain with the
largest influence occurring at very short lengths and with non-zero
airflow. However, over such lengths contact and droplet transmission
are more likely to drive disease transmission.

From the preceding discussion on droplet timescales it is evident
that, in a ventilated environment, droplets are rapidly transported
out of the domain and disease transmission is comparatively too slow
to result in substantial infection. For larger domains convection
times can be significantly increased and the transmission of
infection will become more efficient (e.g. $\tau_c\approx 3.5$ days
for $l=60,000$ m). We estimate $R_{0}=1.34$ and $\lambda=0.005$ for
$d_{1}=4\ \mu$m and $R_{0}=0.0019$ and $\lambda=0.022$ for
$d_{2}=0.4\ \mu$m. The lower value of $R_{0}$ for $d=d_{2}$ is the
result of the small viral load of the droplet which determines the
magnitude of the transmission rate per droplet $\beta_{d}$, Table
\ref{table:Parameters}. Thus, the larger generation rates associated with
smaller droplets do not result in greater transmission.

\subsection{Droplet delay equation}
For a ventilated environment we can neglect droplet diffusion
and Eq.~(\ref{Dropleteqn}) becomes a first order equation. We consider the solution for the particular initial condition $D(x,0)=0$. The characteristic curves satisfy
\begin{displaymath}
\frac{dt}{dx}=\frac{1}{\nu} \Comma
\end{displaymath}
which yields
\begin{equation}\label{A2eqn1}
t=\frac{(x-x_{*})}{\nu} \Comma
\end{equation}
for any $x=x_{*}$ at $t=0$. Along each characteristic curve $D$ will
satisfy
\begin{displaymath}
\frac{dD}{dx}+\frac{1}{\nu}D=\frac{1}{\nu}I(x,\scriptstyle{\frac{x-x_{*}}{\nu}}\displaystyle{)} \Comma
\end{displaymath}
This first order linear equation can be easily solved for $D(x)$ to
obtain
\begin{displaymath}
D(x)=\frac{e^{-\frac{x}{\nu}}}{\nu}\left(
\int_{0}^{x}{e^{\frac{\xi}{\nu}}I(\xi,\scriptstyle{\frac{\xi-x_{*}}{\nu}}\displaystyle{)}}\
d\xi-\int_{0}^{x_{*}}{e^{\frac{\xi}{\nu}}I(\xi,\scriptstyle{\frac{\xi-x_{*}}{\nu}}\displaystyle{)}}\
d\xi \right) \Teleia
\end{displaymath}
The parameter $x_{*}$ can be eliminated using Eq.~(\ref{A2eqn1}) yielding
\begin{displaymath}
D(x,t)=\frac{1}{\nu}\int_{x-\nu
t}^{x}{e^{-\frac{x-\xi}{\nu}}I(\xi,t-\scriptstyle{\frac{x-\xi}{\nu}}\displaystyle{)}}\
d\xi \Teleia
\end{displaymath}
This solution is only valid far from the boundaries, for $\nu t<x<1$, however, it provides an intuitive idea of the droplet dynamics. The role of droplets in disease transmission is thus to introduce a delay into the classic $SIR$ system. Disease transmission at time $t$ depends on the density of infected individuals at a previous time $t-\frac{(x-\xi)}{\nu}$.
The integral also describes how the convection of droplets influences the spatial spread of the disease. It describes the force of infection that individuals at position $\xi$ exert on those at $x$ and as $|x-\xi|\rightarrow\infty$ the droplet density vanishes and no infection will occur. Thus, the transmission risk decays with distance from the source of infection. A corresponding equation can be derived for the diffusion-only scenario showing similar behavior and its derivation is outlined in Appendix A.

\subsection{\label{stability} Stability of a homogeneous population density}
We derive stability criteria for the homogeneous case where the total population density is
uniformly distributed throughout the domain and $N=1$. For the basic disease-free state $(S,I,D)=(1,0,0)$, the evolution of small perturbations $(\hat{S},\hat{I},\hat{D})$ are governed by the linearised equations
\begin{align*}
\hat{S}_{t}&=-\lambda R_{0}\hat{D}+\eta_{p} \hat{S}_{xx},\\
\hat{I}_{t}&=\lambda R_{0}\hat{D}-\lambda \hat{I}+\eta_{p} \hat{I}_{xx},\\
\hat{D}_{t}&=\hat{I}-\hat{D}-\nu\hat{D}_{x}+\eta_{d} \hat{D}_{xx}.
\end{align*}
We consider perturbations proportional to $e^{ikx+\omega t}$, where $\omega$ and $k$ are the frequency and wavenumber, respectively. On substitution into the linearised system we obtain a quadratic equation for $\omega$
\begin{displaymath}
\omega^{2}+\omega(i\nu k+M+ L)-\lambda R_{0}+M L +i\nu k M=0 \Comma
\end{displaymath}
where, for convenience, we define
\begin{displaymath}
M=\lambda+\eta_{p}k^{2},\quad L =1+\eta_{d}k^{2} \Teleia
\end{displaymath}
This has solution
\begin{equation}\label{omega}
\omega(k)=-i\frac{\nu k}{2}-\frac{(M+ L )}{2}\left[1\pm\sqrt{1+\frac{4\lambda R_{0}-4ML-\nu^{2}k^{2}}{(M+L)^{2}}+2i\frac{\nu k(L-M)}{(L+M)^{2}}}\right] \Teleia
\end{equation}
The disease-free state will be unstable when the real part $\Re\{\omega(k)\}>0$ and
disturbances with wavenumber $k$ will grow. Defining the square root in Eq.~(\ref{omega}) as
$p+iq$, with $p>0$ and both $p$ and $q$ real, we can write
\begin{displaymath}
\Re\{\omega\}=-\frac{(M+L)}{2}(1\pm p) \Comma
\end{displaymath}
where for instability we require $p>1$. Now, equating real and imaginary parts yields
\begin{displaymath}
p^{2}-q^{2}=1+\frac{4\lambda R_{0}-4ML-\nu^{2} k^{2}}{(M+L)^{2}},\quad\quad
q=\frac{\nu k(L-M)}{p(L+M)^{2}} \Teleia
\end{displaymath}
\begin{figure}[tp]
\centering
\resizebox{12cm}{9cm}{\includegraphics{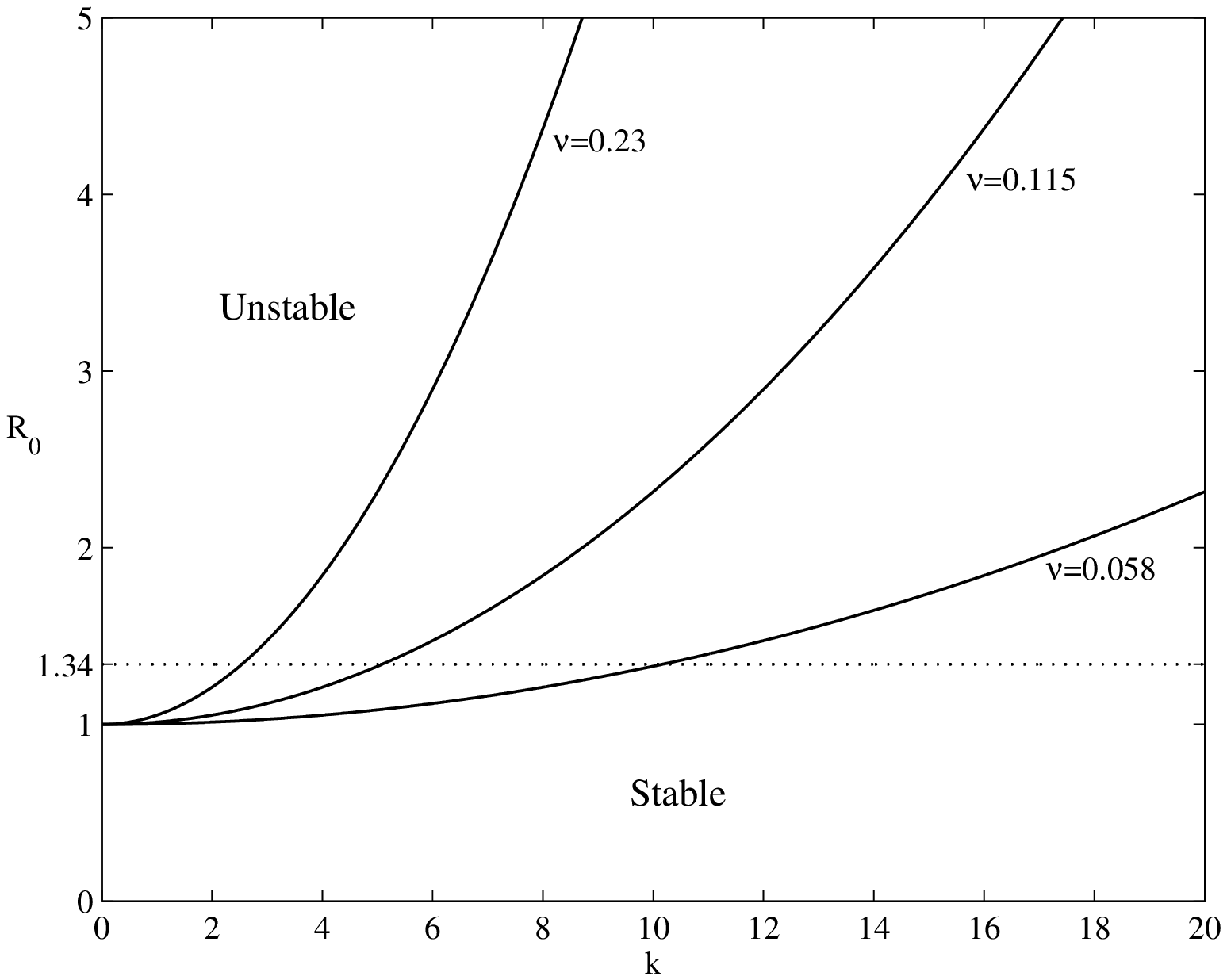}}
\caption{}
\label{stability-curve}
\end{figure}
\noindent Solving for $q$ in terms of $p$ we obtain
\begin{displaymath}
F(p)=p^{2}-\frac{\nu^{2} k^{2}(L-M)^{2}}{p^{2}(L+M)^{4}}=\frac{(M+L)^{2}+4\lambda R_{0}-4ML-\nu^{2}k^{2}}{(M+L)^{2}} \Teleia
\end{displaymath}
Since $F(p)$ is a monotonic increasing function of $p$ the condition
for instability can be expressed as $F(p)>F(1)$ which yields
\begin{equation}\label{instability}
R_{0}>\frac{M L}{\lambda}\left[1+\frac{\nu^{2}k^{2}}{(M+L)^{2}}\right] \Teleia
\end{equation}
We calculate $\eta_{p}=5.77\times 10^{-9}$ and $\eta_{d}=3.57\times
10^{-15}$ and, since $\eta_{d}\ll\eta_{p}\ll1$, we can approximate
Eq.~(\ref{instability}) by
\begin{displaymath}
R_{0}>1+\nu^{2}\frac{k^{2}}{(\lambda+1)^{2}} \Teleia
\end{displaymath}
Clearly, $R_{0}>1$ is a necessary condition for instability and it
follows that the infection will always die out for $d=d_{2}$,
regardless of whether ventilation is present or not, since $R_{0}=0.0019\ll1$. In an
unventilated environment, $\nu=0$, the larger airborne droplets will be transported
by diffusion alone and the uniform state is unstable at all
wavenumbers when $R_{0}>1$ and the infection will spread. In a
ventilated environment, $\nu>0$, disturbances with a wavenumber
satisfying
\begin{equation}\label{wavenumber}
k<\frac{(\lambda+1)}{\nu}\sqrt{R_{0}-1}=k_{\textrm{crit}} \Teleia
\end{equation}
will be unstable. This result shows that the basic homogeneous disease-free state is stable to short wavelength perturbations ($k\to\infty$) and only becomes unstable at long wavelengths ($k\to0$). The stability diagram is sketched in Figure \ref{stability-curve}. Clearly, $R_0$ must exceed unity for the onset of instability and, for fixed $R_0>1$, the critical wavenumber decreases with increasing $\nu$, which physically corresponds to increasing the ventilation velocity and thereby reducing the time it takes for droplets to traverse the domain. Therefore, increasing $\nu$ has a stabilizing effect, by reducing the number of unstable modes that will be amplified.

For a velocity of $v=0.2$ m s$^{-1}$, the critical wavenumber is $k_\textrm{crit}=2.54$ and perturbations with wavelength satisfying $\Lambda>\Lambda_\textrm{crit}=\frac{2\pi}{k}\approx2.47$ will be amplified. Significantly, all permissible wavelengths ($\Lambda<1$) for our domain $x\in(0,1)$ are stable and therefore an arbitrary perturbation to the disease-free state will decay. Reducing the ventilation velocity decreases $\Lambda_\textrm{crit}$ and unstable modes become permissible within the domain and any arbitrary perturbation will grow in time. The analysis indicates that, for a fixed size domain, increasing the ventilation velocity can mitigate the effects of a respiratory disease.

\section{\label{sec6}Results}
We examined the spatial dynamics of the model for droplet size $d_1=4\ \mu$m considering both homogeneous
and heterogeneous initial population-density distributions. The system (\ref{Susceptibleeqn})-(\ref{Dropleteqn}) was solved using a fully implicit finite difference scheme implemented in Matlab$^\circledR$.  

\subsection{\label{unventilated}An unventilated environment}
\begin{figure}[tp]
\centering
\resizebox{17.5cm}{7.5cm}{\includegraphics{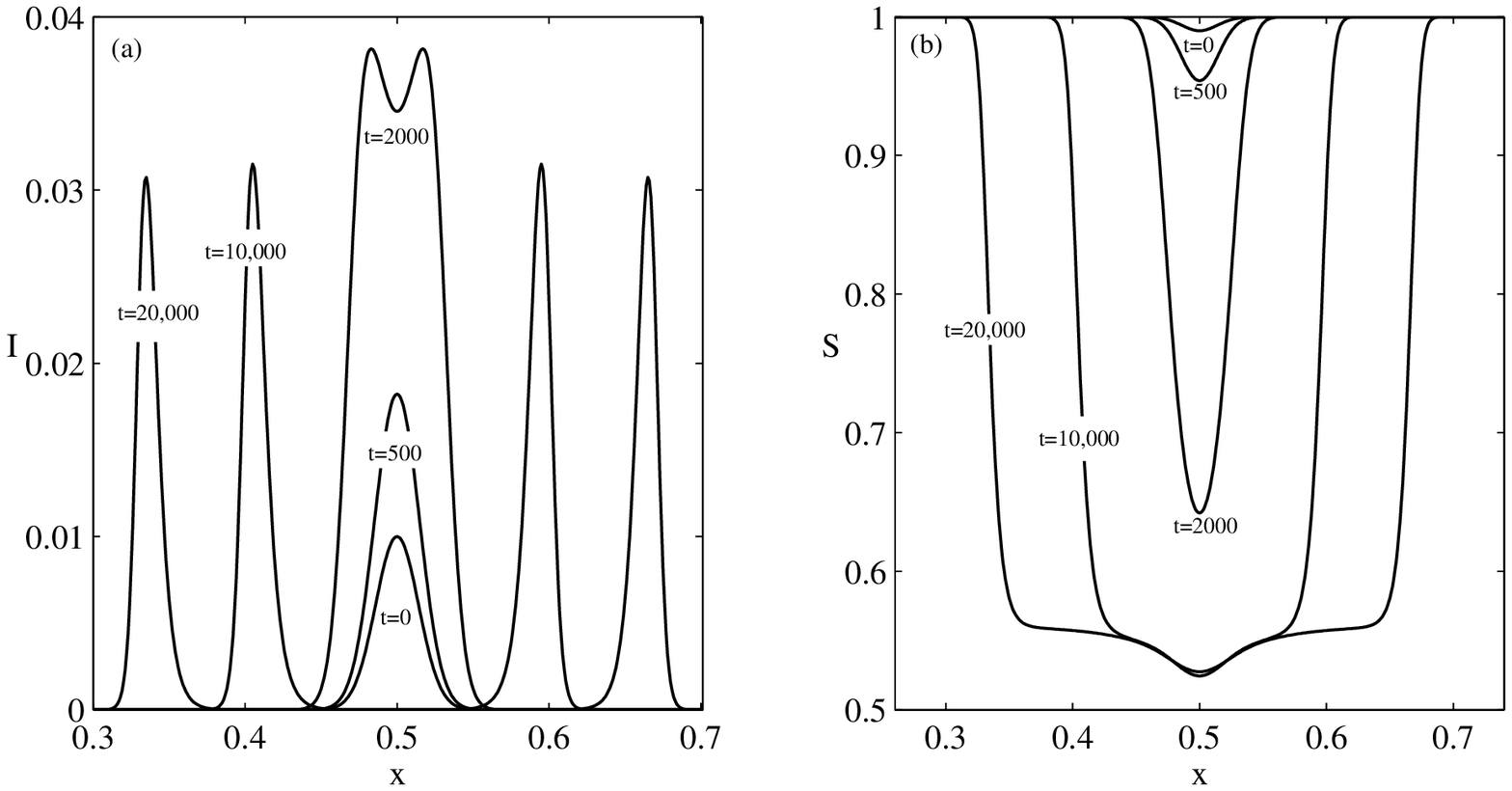}}
\caption{}
\label{fig:DiffusionHomo}
\end{figure}

\begin{figure}[tp]
\centering
\resizebox{10cm}{8cm}{\includegraphics{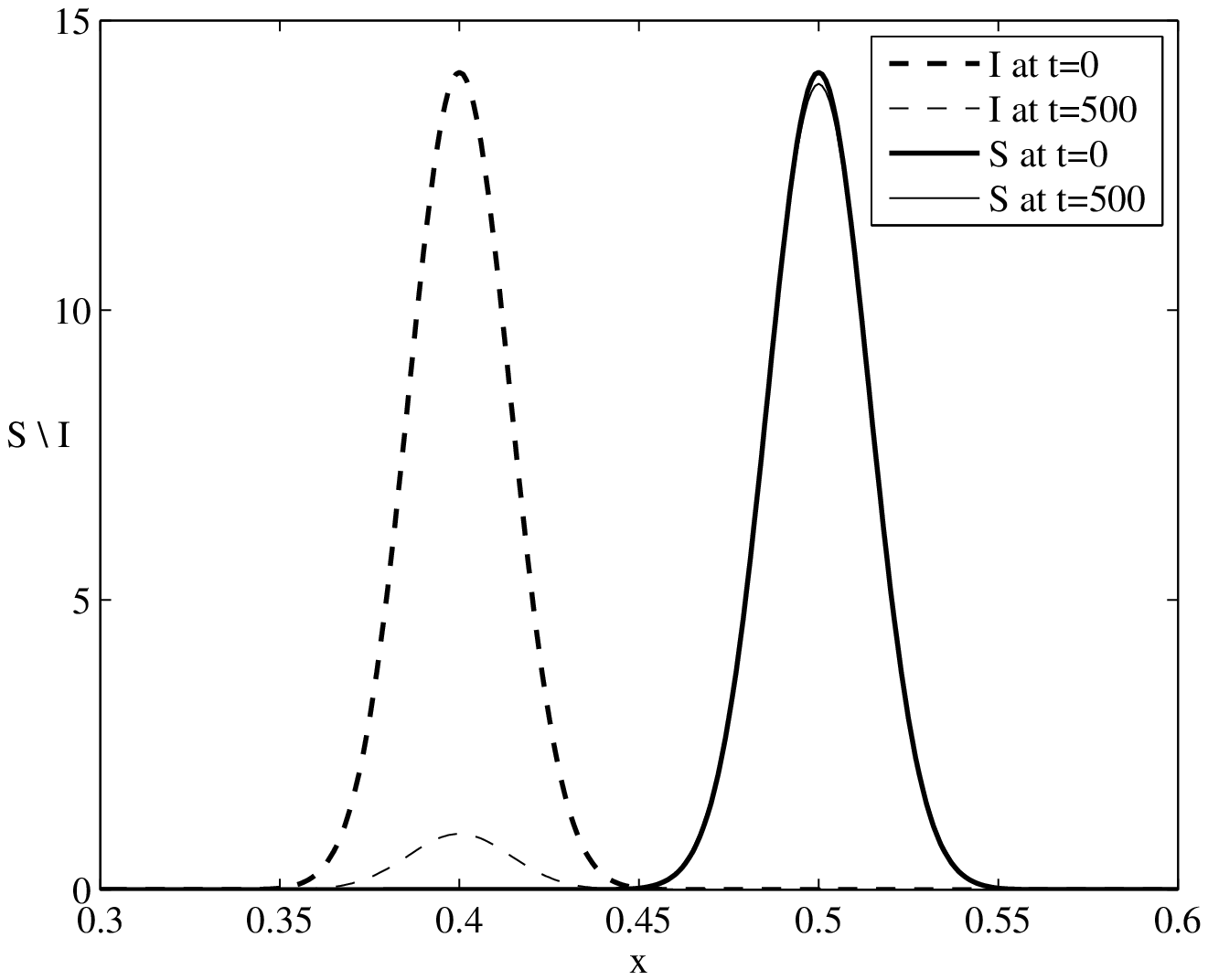}}
\caption{}
\label{fig:DiffusionHetero}
\end{figure}

In an unventilated environment, with $\nu=0$, the spatial spread of
disease will solely depend on the diffusion of people or droplets.
Droplet diffusion is a slow process, for example a $4\ \mu$m droplet
will diffuse approximately $5.7$ mm while airborne. In the same time
a person could diffuse over $7$ meters. Therefore, the spatial
spread of disease will be driven by the diffusion of people as
droplets are essentially stationary relative to the human
population. Mathematically, this observation follows immediately from the
inequality $\eta_{d}\ll\eta_{p}$.
We first consider the disease spread when a localized
infected density is introduced to a uniformly distributed
susceptible population. We find that, for small times ($t\lesssim 5\approx 3\textrm{ hours}$), droplet
density rapidly increases until $D(x,t)\approx I(x,t)$ and
subsequently droplets and infected dynamics are indistinguishable.
This occurs once droplet generation and removal processes are
essentially balanced. This indicates that there is no separation between droplets and infected individuals and close contact is required for efficient transmission. At larger times the localized infectious peak is observed to grow in amplitude, Figure~\ref{fig:DiffusionHomo}(a). An outbreak in a closed population could be expected to persist for many weeks. However, times in excess of $t=2000\approx 53$ days are unrealistic for an influenza epidemic. At such times the model predicts the formation of two infectious traveling pulses, propagating in opposite directions. Accordingly, the density of susceptibles evolves into a wave front slowly infiltrating the completely susceptible population ahead of the front, Figure~\ref{fig:DiffusionHomo}(b). There are several reasons why this behavior is observed. Firstly, the idealized case of a homogeneous population, where the infected population is surrounded by a constant source of susceptibles is uncharacteristic of a true human population distribution. In the absence of susceptible individuals the density of infected individuals would rapidly decay, Figure \ref{fig:DiffusionHetero}. Secondly, once an outbreak is identified within the closed population, interventions such as isolation, quarantine or treatment would be imposed to limit the impact of the outbreak. Diffusion models used to describe the larger scale geographic spread of disease display such traveling wave behavior, with transportation networks simulated using larger diffusion coefficients and timescales of the order of years being considered \cite{noble,murraya}.

Disease spread resulting from a
heterogeneous initial distribution, where localized densities of infected and
susceptible individuals are placed in different regions of the domain, is considered in
Figure \ref{fig:DiffusionHetero}. As before, the droplet density rapidly approaches the
density of infected individuals and the two densities are thereafter
indistinguishable from each other. We find that, even with very
close contact between the two groups, diffusion is too slow a
process to effectively transmit infection. In the absence of a
susceptible population, and since $\eta_{p}\ll1$, the density of infectious individuals at any point $x=x_{0}$ will decay via
$I(t)=I_0(x_{0})e^{-\lambda t}\rightarrow 0$ as $t\rightarrow\infty$.
The lifetime of the infected population is thus
$\frac{1}{\lambda}$ which (dimensionally) corresponds to the disease infectivity time scale $\tau_i$ of 5 days. The infected population recovers before diffusion has time to effectively transmit the infection, $\tau_i\ll\tau_p$. To conclude, in the absence of an ambient airflow, diffusion is an inefficient mechanism for the spatial spread of disease which will, presumably, be driven by contact and droplet transmission during close contact events.

\subsection{\label{ventilated}A ventilated environment}
\begin{figure}[tp]
\centering
\resizebox{17.5cm}{7.5cm}{\includegraphics{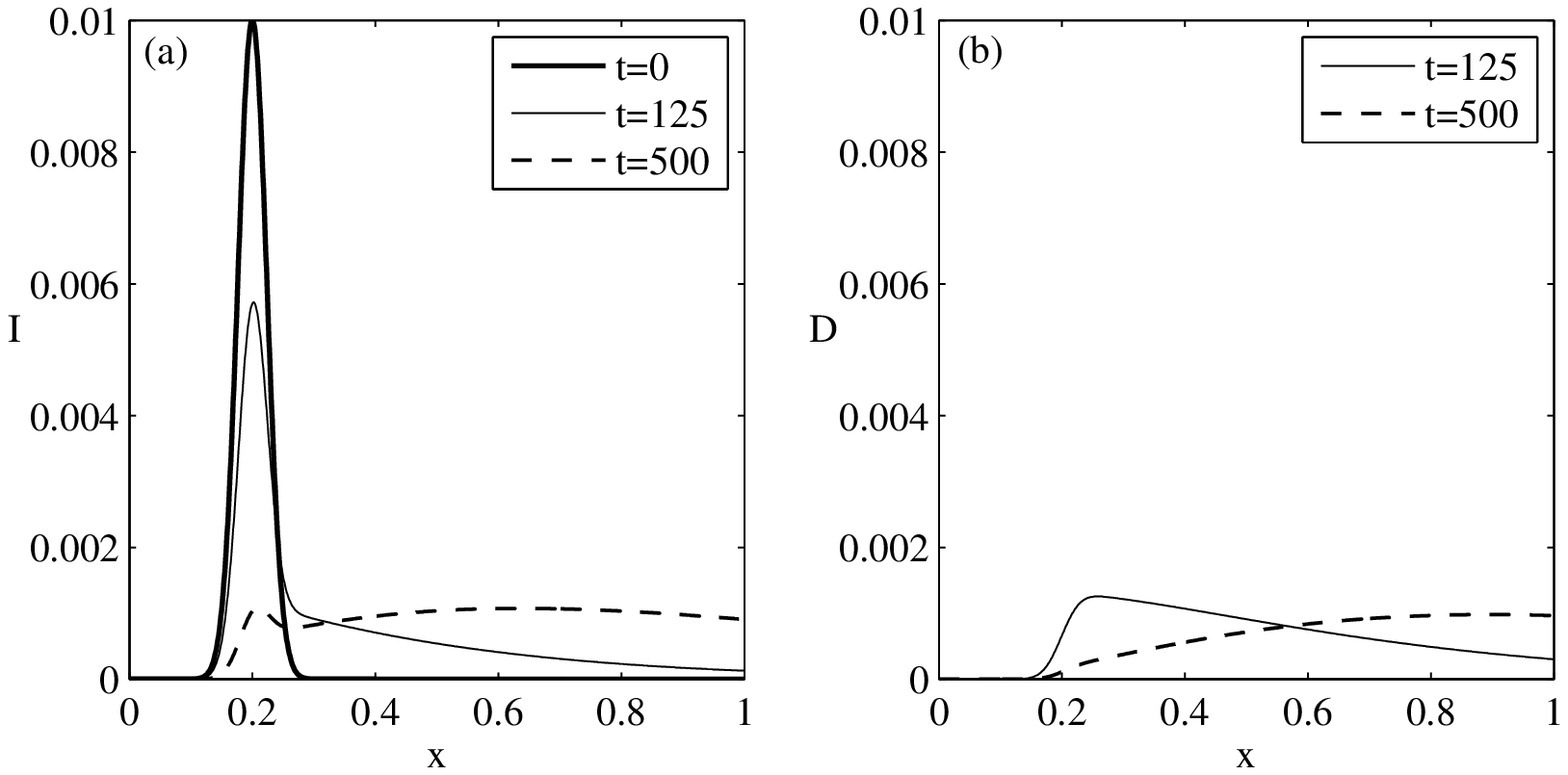}}
\caption{}
\label{fig:convectionstable}
\end{figure}

In a ventilated environment, with $\nu>0$, the infection dynamics are determined by the droplet size, the ambient air velocity and
the length of the spatial domain. For a fixed droplet size and domain length, the wavenumbers that satisfy Eq.~(\ref{wavenumber}) are determined solely by the ventilation velocity $v$. As discussed in Section \ref{stability}, when $v=0.2$ m s$^{-1}$, all permissible wavenumbers are stable. The evolution of an arbitrary perturbation under such conditions is shown in Figure \ref{fig:convectionstable}. Droplets generated by the infectious population are rapidly transported out of the domain causing minimal infection as they travel through the susceptible population, since the convective timescale is much less than that required for transmission $\tau_c\ll\tau_t$, and the infectious curve decays.

\begin{figure}[tp]
\centering
\resizebox{17.5cm}{7.5cm}{\includegraphics{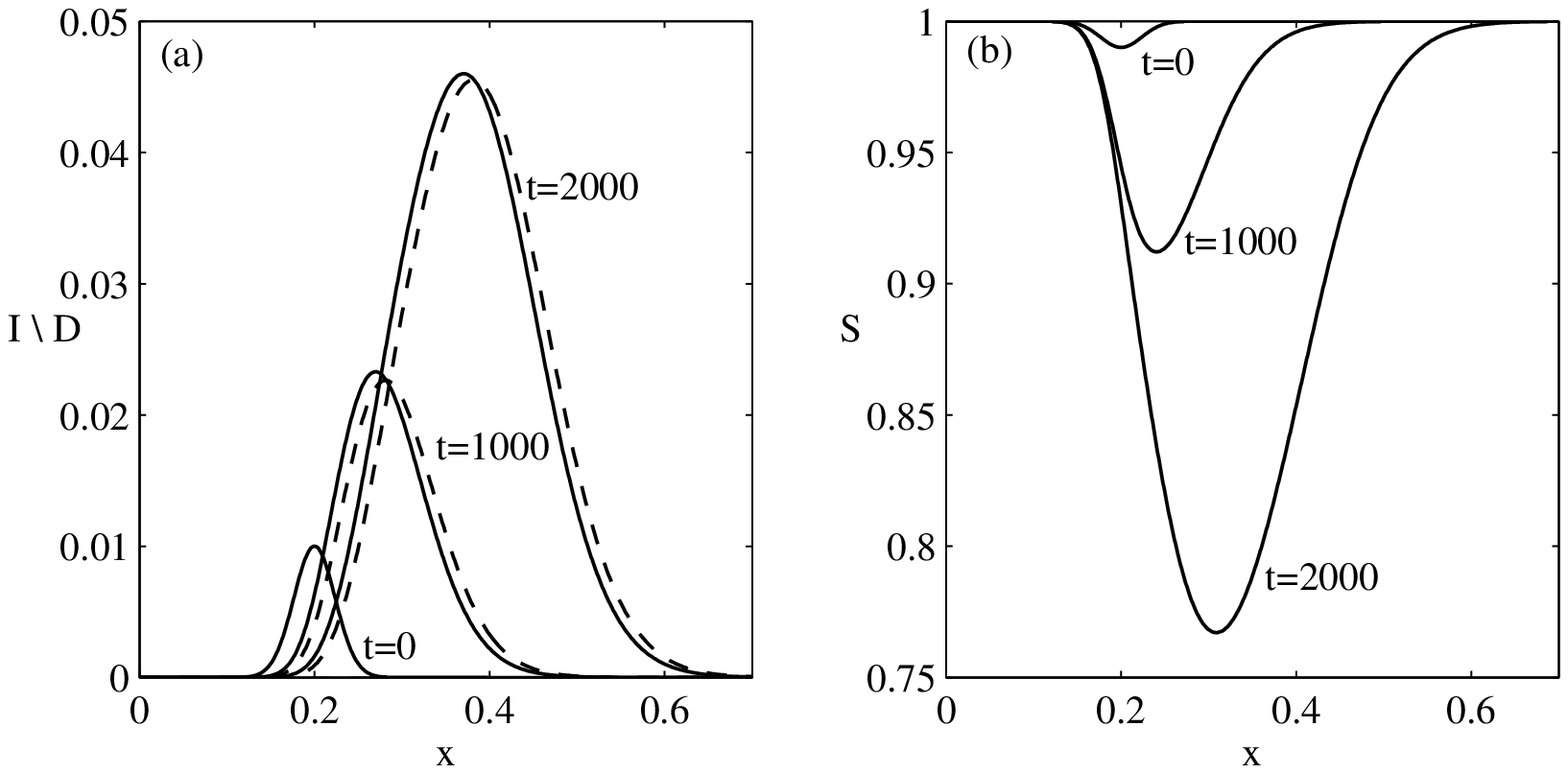}}
\caption{}
\label{fig:convectionunstable}
\end{figure}

If we rewrite (\ref{wavenumber}), we find that the basic-state will be stable to an arbitrary perturbation provided the ventilation velocity satisfies
\begin{displaymath}
v>\frac{\alpha_d l(\lambda+1)}{2\pi}\sqrt{R_0-1}=v_\textrm{crit}.
\end{displaymath}
For parameter values listed in Table \ref{table:Parameters} we find that $v_\textrm{crit}=0.08$ m s$^{-1}$ and ventilation velocities of this magnitude and above are sufficient to prevent the spatial spread by the airborne route for a uniformly distributed population. However, we emphasize that this is an approximate value obtained for the relatively small diffusion of people and neglecting the possibility of contact and/or droplet transmission. In addition, the true air velocity will depend on the distance from the ventilation source, with larger velocities closer to the source. The unstable evolution of an arbitrary perturbation when $v=0.01$ m s$^{-1}$ is shown in Figure \ref{fig:convectionunstable}. An infectious pulse can be seen to propagate through the susceptible population in the direction of positive air flow. Droplet and infected densities are qualitatively similar and only slightly out of phase with each other, with droplets propagating ahead of the infectious pulse. Dynamically this implies that, for small air velocities, the dynamics of droplets and infectious individuals are closely coupled and only separate from each other at sufficiently high velocities.\\

\begin{figure}[htp]
\centering
\resizebox{16cm}{18cm}{\includegraphics{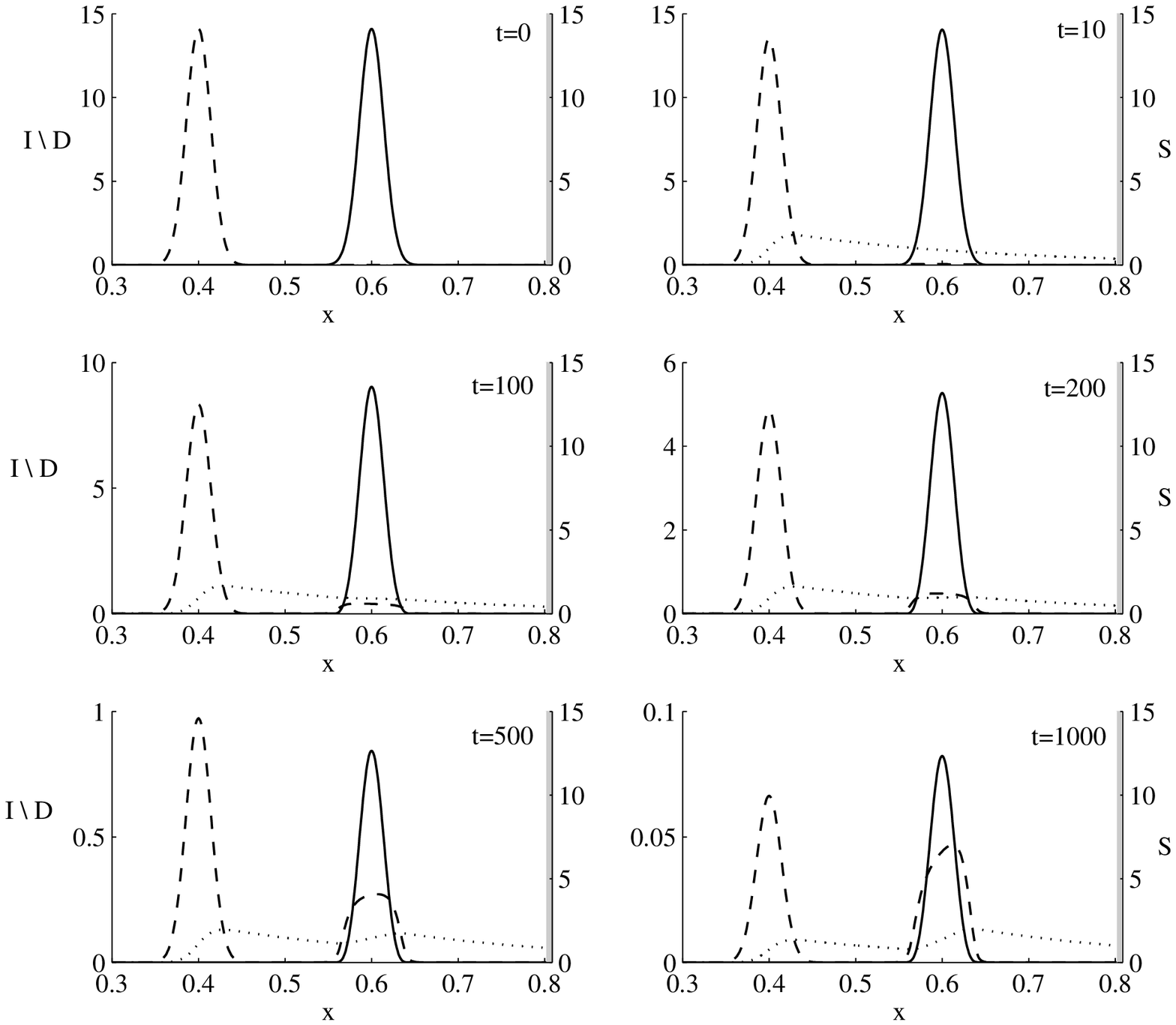}}
\caption{}
\label{fig:convectionhetero}
\end{figure}

Finally, we consider the spatial transmission of infection resulting from a heterogeneous
initial population distribution, where infected and susceptible populations are placed at separate locations in the domain, Figure~\ref{fig:convectionhetero}. For small times, $t\lesssim 10\approx 6$ hours, the infected population generate droplets which are subsequently transported in the direction of positive air flow. A secondary curve for infectious individuals forms when droplets encounter the susceptible population, $t=100\approx 2.5$ days. At later times these secondary cases produce droplets of their own and a secondary peak in the droplet curve can be observed, $t=500\approx 13$ days. The amplitude of the secondary curve for infectious individuals is obviously influenced by the initial distance between the infectious and susceptible populations. At large times, $t=1000\approx 26$ days, it is clear that the amplitudes of both curves for infectious individuals decay. This behaviour results from the decay of the initial infectious population through recovery and the subsequent reduction in droplets being convected towards the susceptible region. This demonstrates that it is possible for infection to occur in the absence of an infected population and purely as the result of the airborne non-diffusive transport of aerosol droplets.

\section{\label{Conclusions}Discussion}
In this work we developed a model for the spatial spread of an airborne infection driven by pathogen-carrying droplets.
Two droplet diameter values, $d_1=4\ \mu$m and $d_2=0.4\ \mu$m, were considered that are representative of experimentally determined droplet size distributions. Droplet dynamics are governed by generation and removal processes, the latter being dominated by gravitational settling and pathogen inactivation. Inactivation is a particularly crucial removal process for smaller droplets that could remain airborne for days. Smaller droplets are found to be a weak disease vector. Their relatively large generation rates do not result in greater transmission due to the small viral load and the associated duration required to transmit infection. Transmission is, thus, dominated by the larger droplets.

A delay equation was derived for the droplet density as a function of the infected population density. The role of droplets in disease transmission is to introduce a delay into the system, with disease transmission at a given time dependent on the number of infected individuals present at a previous time. The equation also highlights how droplet-driven transmission decays with distance from the source of infection.

The relative importance of diffusive and convective processes in the spatial spread of infection was investigated. Two initial population distributions were considered: spatially homogeneous or heterogeneous. In both cases droplet diffusion is shown to be a slow process with disease spread, in an unventilated environment, driven by human movement. This result follows from the observation that the time require for droplets to diffuse is significantly greater than that for humans, $\tau_p\ll\tau_d$. In the homogeneous scenario and for long time scales, the model displays the classic infectious wave propagating through a susceptible population, following an initial transient state until a balance is achieved between droplet generation and removal processes. In contrast, human diffusion is an insufficient mechanism to transmit disease in the heterogeneous scenario, with the infected population recovering before encountering susceptible individuals, $\tau_i\ll\tau_d$. However, the inclusion of ventilation effects and the subsequent transport of droplets from the source of infection can result in a secondary outbreak if susceptible individuals are encountered. This signifies that infection is possible without direct contact between susceptible and infected individuals. Furthermore, it was shown that increasing the velocity above a critical value can impair disease transmission in a homogeneously distributed population as droplets will be rapidly transported out of the domain causing minimal infection since the time required for transmission is large relative to the convective time scale  $\tau_c\ll\tau_t$.

The use of ventilation to prevent disease transmission is well-accepted by both society and science. However, little work has been done on implementing control strategies based on this knowledge. Identifying optimum air velocities for indoor environments could mitigate transmission and reduce the disease burden in health-care facilities, schools and other densely populated locations. The model presented here indicates that, in a fixed size environment, the distribution of sick/healthy individuals and the ambient air velocity are the primary factors to consider when analysing such an intervention.

\renewcommand{\theequation}{A.\arabic{equation}}
\setcounter{equation}{0}
\section*{Appendix A - Solution of droplet equation with diffusion}
\label{Appendix1}
In the absence of ventilation effects the droplet equation (\ref{Dropleteqn}) takes the form
\begin{equation}\label{Deqn}
\frac{\partial D}{\partial t}=-D+I+\eta_{d}\frac{\partial^{2}D}{\partial x^{2}},
\end{equation}
and, for convenience, we consider the solution on an infinite domain as droplet diffusion is a sufficiently slow process that boundaries do not interfere with the solutions, Figures \ref{fig:DiffusionHomo} and \ref{fig:DiffusionHetero}. We let $D(x,0)=\phi(x)$ denote the initial droplet density and assume a solution of the form
\begin{displaymath}
D(x,t)=H(t)G(x,t),
\end{displaymath}
which yields
\begin{displaymath}
\frac{\partial G}{\partial t}+\frac{G(x,t)}{H(t)}\frac{dH}{dt}=-G(x,t)+\frac{I(x,t)}{H(t)}+\eta_{d}\frac{\partial^{2} G}{\partial x^{2}},\quad\textrm{with}\quad H(t)\neq0.
\end{displaymath}
We choose $H(t)$ to satisfy
\begin{displaymath}
\frac{dH}{dt}=-H,\quad\quad \Rightarrow\quad\quad H(t)=e^{-t},
\end{displaymath}
and the function $G(x,t)$ then satisfies the inhomogeneous diffusion equation
\begin{displaymath}
\frac{\partial G}{\partial t}=f(x,t)+\eta_{d}\frac{\partial^{2}G}{\partial x^{2}},
\end{displaymath}
where $f(x,t)=\frac{I(x,t)}{H(t)}$ and with initial condition $G(x,0)=\phi(x)$. Following Kevorkian \cite{kevorkian}, the solution for $G(x,t)$ will be of the form
\begin{displaymath}
G(x,t)=u(x,t)+v(x,t),
\end{displaymath}
where $u$ and $v$ are the solutions of the simplified diffusion equations
\begin{align*}
&u_{t}=\eta_{d}u_{xx}+f(x,t), &\quad &v_{t}=\eta_{d}v_{xx},&\\
&u(x,0)=0,&\quad & v(x,0)=\phi(x).&
\end{align*}
These equations can easily be solved to obtain
\begin{align*}
u(x,t)&=\int_{0}^{t}{\int_{-\infty}^{\infty}{\frac{1}{\sqrt{4\pi \eta_{d}(t-t')}}\exp{\left(-\frac{(x-x')^{2}}{4\eta_{d}(t-t')}\right)}f(x',t')}}\ dx'dt',\\
v(x,t)&=\frac{1}{\sqrt{4\pi \eta_{d}t}}\int_{-\infty}^{\infty}{\exp{\left(-\frac{(x-x')^{2}}{4\eta_{d}t}\right)}\phi(x')}\ dx',
\end{align*}
and the solution of (\ref{Deqn}) is then
\begin{align*}
D(x,t)=&e^{-t}\int_{0}^{t}{\int_{-\infty}^{\infty}{\frac{1}{\sqrt{4\pi \eta_{d}(t-t')}}\exp{\left(-\frac{(x-x')^{2}}{4\eta_{d}(t-t')}\right)}\frac{I(x',t')}{H(t')}}}\ dx'dt'\\
&+e^{-t}\frac{1}{\sqrt{4\pi \eta_{d}t}}\int_{-\infty}^{\infty}{\exp{\left(-\frac{(x-x')^{2}}{4\eta_{d}t}\right)}\phi(x')}\ dx'.
\end{align*}
\bigskip



\roman{table}
\setcounter{equation}{0}  

\newpage
\btab[htp]
\begin{center}
\begin{tabular}{lll} \hline
time scale (s)& & dimensionless parameter  \\ \hline
$\tau_i = \frac{1}{\mu_i}$ & disease infectivity &  \\
$\tau_t = \frac{\alpha_d}{\beta_d \kappa_d}$ & disease transmission & $R_0 \equiv \frac{\beta_{d}\kappa_{d}}{\alpha_{d}\mu_i} = \frac{\tau_i}{\tau_t} $ \\
$\tau_r = \frac{1}{\alpha_d}$ & droplet removal (droplet infectivity) & $\lambda=\frac{\mu_{i}}{\alpha_{d}} = \frac{\tau_r}{\tau_i}$ \\
$\tau_c = \frac{l}{v}$ & convective & $\nu=\frac{v}{\alpha_{d} l} = \frac{\tau_r}{\tau_c} $ \\
$\tau_{d,p} = \frac{l^2}{D_{d,p}}$ & diffusion (droplet, person)
& $\eta_{d,p}=\frac{D_{d,p}}{\alpha_{d} l^{2}} = \frac{\tau_r}{\tau_{d,p}}$ \\
\hline
\end{tabular}
\end{center}
\caption{Characteristic time scales and derived dimensionless parameters.}
\label{table:Scales}
\etab

\newpage
\btab[htp]
\begin{center}
\selectfont{\footnotesize{\begin{tabular}{ l l l l l} \hline
\multicolumn{2}{l}{parameter}  & \multicolumn{2}{l}{value} & \\ \hline
$\mu_i$ & infection recovery rate & \multicolumn{2}{l}{0.2 per day}   &   \\
$c$ & contact rate & \multicolumn{2}{l}{13 per day}&   \\
$\rho_{p}$ & pathogen concentration in the lung fluid & \multicolumn{2}{l}{$3.71\times10^{6}$ pathogens cm$^{-3}$} &   \\
$B$ & breathing rate & \multicolumn{2}{l}{$24$ m$^{3}$ d$^{-1}$} &  \\
$V_{cl}$ & personal-cloud volume of an infected person & \multicolumn{2}{l}{$8$ m$^{3}$} &  \\
$p_{d}$ & infection probability by an inhaled pathogen & \multicolumn{2}{l}{0.052} &   \\
$\tau_{ct}$ & characteristic breathing (contact) time & \multicolumn{2}{l}{20 min} &   \\
$\beta_{p}$ & transmission rate per inhaled pathogen & \multicolumn{2}{l}{0.028 per day} &  \\
$v$ & air velocity & \multicolumn{2}{l}{$0.2$ m s$^{-1}$} &  \\
$D^{tur}_{d}$ & turbulent diffusivity of droplet & \multicolumn{2}{l}{$10^{-3}$ m$^{2}$ s$^{-1}$}&  \\
$D_{p}$ & diffusivity of people & \multicolumn{2}{l}{$10^{-5}$ m$^{2}$ s$^{-1}$} &  \\
$\mu_p$ & pathogen inactivation rate & \multicolumn{2}{l}{$8.64$ per day} & \\
$l$ & domain length & \multicolumn{2}{l}{$2000$ m} & \\
\multicolumn{5}{l}{\textit{Parameters dependent on droplet size}} \\
$d$ & droplet diameter (post-evaporation) & \multicolumn{2}{l}{$d_{1}$=4$ \mu$m} &  \\
&  & \multicolumn{2}{l}{$d_{2}$=0.4$ \mu$m} &  \\
$V_d$ & pre-evaporation (spherical) droplet volume & \multicolumn{2}{l}{$V_{d_{1}}=2.68\times 10^{-10}$ cm$^{3}$} &  \\
&  & \multicolumn{2}{l}{$V_{d_{2}}=2.68\times 10^{-13}$ cm$^{3}$} &  \\
$N_p^0$ & initial number of pathogens per droplet & \multicolumn{2}{l}{$N_{p}(d_{1})=9.95\times 10^{-4}$} &  \\
&  & \multicolumn{2}{l}{$N_{p}(d_{2})=9.95\times 10^{-7}$} &  \\
$q_{d}$ & inhaled droplet deposition probability & \multicolumn{2}{l}{$q_{d_{1}}=0.88$} &  \\
&  & \multicolumn{2}{l}{$q_{d_{2}}=0.2$} &  \\
$\beta_{d}$ & transmission rate per inhaled droplet & \multicolumn{2}{l}{$\beta_{d_{1}}=2.45\times 10^{-5}$ per day} &  \\
&  & \multicolumn{2}{l}{$\beta_{d_{2}}=5.57\times 10^{-9}$ per day} &  \\
$\kappa_{d}$ & droplet production rate & \multicolumn{2}{l}{$\kappa_{d_{1}}=4.1\times 10^{5}$ per day} & \\
&  & \multicolumn{2}{l}{$\kappa_{d_{2}}=6.14\times 10^{5}$ per day} &  \\
$\theta_{d}$ & gravitational settling rate & \multicolumn{2}{l}{$\theta_{d_{1}}=28.80$ per day }  & \\
&  & \multicolumn{2}{l}{$\theta_{d_{2}}=0.39$ per day } &  \\
$\alpha_{d}$ & droplet removal rate  & \multicolumn{2}{l}{$\alpha_{d_{1}}=37.44$ per day} & \\
&  & \multicolumn{2}{l}{$\alpha_{d_{2}}=9.03$ per day} &  \\
$D_{d}$ & molecular diffusivity of droplet & \multicolumn{2}{l}{$D_{d_{1}}=6.2\times10^{-12}$ m$^{2}$ s$^{-1}$}  &  \\
& & \multicolumn{2}{l}{$D_{d_{2}}=8.34\times10^{-11}$ m$^{2}$ s$^{-1}$}& \\ \hline
\end{tabular}}}
\end{center}
\caption{Parameter values}
\label{table:Parameters}
\etab

\newpage
\section*{Figure captions}
\begin{description}

\item[Figure 1:] Stability diagram for a homogeneously distributed population.

\item[Figure 2:] Model dynamics of airborne-influenza transmission driven by the
diffusion of people, with droplet diameter $d_1=4 \mu$m. Initial conditions are $I_0(x)=0.01 e^{-k^{2}(x-0.5)^{2}}$,
$S_0(x)=1-I_0(x)$ with wavenumber $k=50$.
(a): Density of infected individuals. (b): Density of susceptible individuals. All variables are dimensionless and scaled following (\ref{scales}).

\item[Figure 3:] Model dynamics of influenza transmission driven by the
diffusion of people, with droplet diameter $d_{1}=4 \mu$m. The graph
shows the densities of susceptible (solid line) and infected (dashed line)
individuals with initial
condition $I_0(x)=\frac{25}{\sqrt{\pi}}e^{-k^{2}(x-0.4)^{2}}$,
$S_0(x)=\frac{25}{\sqrt{\pi}}e^{-k^{2}(x-0.5)^{2}}$ and
wavenumber $k=50$. All variables are dimensionless and scaled following (\ref{scales}).

\item[Figure 4:] Model dynamics of an influenza outbreak driven by the
convection of droplets, with respirable droplet diameter $d_1=4 \mu$m and
ventilation velocity $v=0.2$ m s$^{-1}$. Initial conditions are $I_0(x)=0.01 e^{-k^{2}(x-0.2)^{2}}$,
$S_0(x)=1-I_0(x)$ with $k=30$. (a): Density of infected individuals (b): Density of droplets. All variables are dimensionless and scaled following (\ref{scales}).

\item[Figure 5:] Model dynamics of an influenza outbreak driven by the
convection of droplets, with respirable droplet diameter $d_1=4 \mu$m and
ventilation velocity $v=0.01$ m s$^{-1}$. Initial conditions are $I_0(x)=0.01 e^{-k^{2}(x-0.2)^{2}}$,
$S_0(x)=1-I_0(x)$ with $k=30$. (a): Densities of infected individuals (solid line)
and droplets (dashed line). (b): Density of susceptible individuals. All variables are dimensionless and scaled following (\ref{scales}).

\item[Figure 6:] Model dynamics of an influenza outbreak driven by the
convection of droplets, with respirable droplet diameter $d_1 = 4 \mu$m and
ventilation velocity $v=0.2$ m s$^{-1}$. The graphs show the
density of susceptibles (solid line) on the right vertical axis. The densities of infected individuals (dashed line) and
droplets (dotted line) are shown on the left vertical axis.
Initial conditions are $I_0(x)=\frac{25}{\sqrt{\pi}}e^{-k^{2}(x-0.4)^{2}}$,
$S_0(x)=\frac{25}{\sqrt{\pi}}e^{-k^{2}(x-0.6)^{2}}$ and with $k=50$. All variables are dimensionless and scaled following (\ref{scales}).

\end{description}

\end{document}